\documentclass[letterpaper,useAMS,usenatbib]{mn2e}
\usepackage{xspace}
\usepackage{graphicx}
\usepackage{hyperref}
\usepackage{verbatim}
\usepackage{Times}
\usepackage{bm}
\usepackage[T1]{fontenc} 
\usepackage{aecompl}

\newcommand{\mnras}{MNRAS}

\newcommand{\be}{\begin{equation}}
\newcommand{\ee}{\end{equation}}
\newcommand{\bea}{\begin{eqnarray}}
\newcommand{\eea}{\end{eqnarray}}

\def\mnras{MNRAS}

\newcommand{\bs}{\boldsymbol}

\def\hmass{M^*_{{\rm h},j}}

\def\numobs{N^{\rm obs}}

\def\data{\bs{d}}
\def\allDat{\bs{D}}

\def\z8{$z_{850}$}
\def\z{z$_{850}$}

\def\hdat{\bs{h}}
\def\numhosts{N_{\rm h}}

\def\prob{{\rm Pr}}

\def\be{\begin{equation}}
\def\ee{\end{equation}}

\def\modpars{\bs{\theta}}

\title[Luminous satellites of gravitational lenses]{Do gravitational lens galaxies have an excess of luminous substructure?}

\author[Nierenberg et al.]{
A.~M.~Nierenberg$^{1,2}$, 
D.~Oldenburg$^{1}$, 
T.~Treu$^{1,3}$
\medskip\\
$^1$Department of Physics, University of California, Santa Barbara, CA 93106, USA\\
$^2$amn01@physics.ucsb.edu\\
$^3$ Packard Fellow \\ 
}

\begin{document}
\date{Accepted for publication in \mnras}
\pagerange{\pageref{firstpage}--\pageref{lastpage}}\pubyear{2013}

\maketitle           

\label{firstpage}

\begin{abstract}
Strong gravitational lensing can be used to directly measure the mass function of their satellites, thus testing one of the fundamental predictions of cold dark matter cosmological models. Given the importance of this test it is essential to ensure that galaxies acting as strong lenses have dark and luminous satellites which are representative of the overall galaxy population. We address this issue by measuring the number and spatial distribution of luminous satellites in ACS imaging around lens galaxies from the Sloan Lens Advanced Camera for Surveys (SLACS) lenses, and comparing them with the satellite population in ACS imaging of non lens galaxies selected from COSMOS, which has similar depth and resolution to the ACS images of SLACS lenses. In order to compare the samples of lens and non lens galaxies, which have intrinsically different stellar mass distributions, we measure, for the first time, the number of satellites per host as a continuous function of host stellar mass for both populations. We find that the number of satellites as a function of host stellar mass, as well as the spatial distribution are consistent between the samples. Using these results, we predict the number of satellites we would expect to find around a subset of the Cosmic Lens All Sky Survey (CLASS) lenses, and find a result consistent with the the number observed by \citet{Jackson++10}. Thus we conclude that within our measurement uncertainties there is no significant difference in the satellite populations of lens and non lens galaxies.

\end{abstract}

\begin{keywords}
dark matter -- 
galaxies: dwarf --
galaxies: haloes --
\end{keywords}
\setcounter{footnote}{1}

\section{Introduction}

A key prediction of $\Lambda$CDM is that galaxy-mass haloes should be surrounded by thousands
of low mass subhaloes. In apparent contrast to this prediction, only tens of companion satellite galaxies have been detected around the Milky Way \citep{Strigari++07}.  This indicates that either $\Lambda$CDM is incorrect and the subhaloes do not exist, or that they do exist and do not form stars or retain sufficient amounts of gas to be detected \citep{Papastergis++11}. Even if future surveys such as LSST are able to detect a new population of ultra faint dwarf galaxies around the Milky Way, (e.g, Tollerud et al. 2008) the detection and study of these satellites outside of the Local Group would be unfeasible with present technology.

Therefore, a measurement of the subhalo mass function which is independent of star formation is necessary to confirm the accuracy of $\Lambda$CDM in this low mass regime. There are several promising avenues for measuring the subhalo mass function in the Local Group, including direct detection of the annihilation of dark matter particles \citep{Strigari++12B}, and disruptions of tidal streams by dark haloes \citep{Carlberg++12}.

 Outside of the Local Group, a powerful way to measure the subhalo mass function without relying on the presence of baryons is with strong gravitational lensing by a galaxy deflector. When a luminous background source is strongly lensed by an intervening galaxy, the number, positions, and magnifications of the images that appear depend only on the mass distribution of the lens galaxy and the relative angular positions of the lens and the source \citep[see][and references therein]{Treu++10}. Substructure along the line of sight can be identified via a deviation in the magnification, positions or time delays of lensed images from what would be expected given a smooth deflector mass distribution \citep[e.g.][]{Mao++98,Metcalf++01,Keeton++09}. 

Gravitational lensing has been applied to a variety of lens systems in order to detect subhaloes and place constraints on their mass function \citep[e.g.][]{Dalal++02, Vegetti++12, Fadley++12,Xu++09, Xu++13}. Due to the small numbers of systems available for this type of analysis, the uncertainties on the inferred subhalo mass function remain large. However, future wide-area surveys such as LSST and PANSTARRS are expected to find thousands of new lens galaxies \citep{Oguri++10} which can be analysed for substructure. Furthermore, the next generation of large telescopes and adaptive optics will make the deep, high resolution imaging required for the analyses of lensed images fast and therefore feasible for a large number of lenses, with sensitivity extending to lower masses. 

In order to use the subhalo mass function measured around gravitational lenses as a test of $\Lambda$CDM, it is crucial to understand the lensing selection function, and in particular whether lens galaxies are representative of field galaxies. Several works have used simulations of different galaxy scale lenses with background point sources to determine which dark matter halo properties, and survey selection criteria affect the gravitational lensing cross section \citep{Keeton++04,Mandelbaum++09,VandeVen++09, Arneson++12,Dobler++08} . These studies found that the lensing cross section is by far most strongly dependent on the surface mass density of the deflector.

\citet{Treu++09}, showed that lens galaxies in the Sloan Lens Advanced Camera for Surveys (SLACS) \citep{Bol++04,Bol++05,Bol++06,Aug++09} exist in environments with densities which are consistent with those of non lens galaxies with similar stellar masses and velocity dispersions. \citet{Fassnacht++11} found a similar result for a sample of intermediate redshift lens galaxies.  Furthermore, \citet{Aug++10} and \citet{Koc++00} showed that the fundamental plane for SLACS and CLASS \citep{Mye++03,Bro++03} lens galaxies is consistent with that of non-lens, early-type galaxies.


In order to generalise the measurements of the subhalo population mass function of lens galaxies it is essential that we understand whether lens galaxies have substructure populations typical of their non-lens counterparts. Outside of the Local Group, the comparisons between the subhaloes of lens and non-lens galaxies are only possible for those subhaloes which contain a significant population of stars. Those subhaloes which contain stars are expected to be the most massive, and therefore affect the lensing cross section most strongly.  We expect that the properties of the subhalo population which significantly alter the probability of lensing should be evident in the subhaloes containing luminous galaxies which are the most massive. Of course, with this sort of analysis we cannot test whether there are indirect selection effects connecting lens galaxies and the dark subhalo population, which do not directly influence the lensing cross-section.

As an important test, \citet{Jackson++10} counted the number of projected companions within 20 kpc apertures around lens galaxies in SLACS and CLASS, and compared them with the number of objects around early-type galaxies selected from Sloan and COSMOS ACS imaging. \citet{Jackson++10} found that SLACS lenses had a similar number of companion objects to other early-type galaxies in Sloan, while CLASS lenses had orders of magnitude more companions in projection than non-lens galaxies in COSMOS field galaxies. 


In this work we revisit the question of the populations of luminous companions of gravitational lenses, taking into account several important developments with regards to the detection and study of satellites. First, we apply the image processing and statistical analysis developed in \citet{Nierenberg++11, Nierenberg++12} (hereafter N11, N12) to ACS images of SLACS lenses, which enables us to directly study the satellite population of these galaxies to more than a thousand times fainter than the host galaxies, and very close in projection ($\sim$1$\farcs 0$). 

 Secondly, we infer, for the first time, the dependence of the number of satellites on the host galaxy stellar mass for both SLACS lenses and COSMOS field galaxies. This allows us to rigorously incorporate differences in the distribution of host stellar masses when comparing between the two samples, which is essential given the strong observed dependence on the number of satellites as a function of host stellar mass \citep[N12,][]{Wang++12}.
  
The paper is organised as follows: in \S \ref{sec:lensSelection} we
describe the selection of the sample of lens and non-lens host galaxies around which we study satellite galaxies. 
In \S \ref{sec:satelliteDetection} we describe our procedure for detecting companion objects around each of the samples of hosts. 
In \S \ref{sec:statistics} we explain the statistical method we use to infer the typical number of satellites per host galaxy
as a function of stellar mass and redshift.  In \S \ref{sec:results} we compare the inferred number of satellites per host
for the SLACS and non-lens galaxies. In \S \ref{sec:CLASScompare} we compare the number of satellites around COSMOS and CLASS hosts. Finally in \S
\ref{sec:Discussion} we conclude with a discussion and summary of the
results. 

\section{Host Galaxy Selection}
\label{sec:lensSelection}
We restrict the galaxies analysed in this work to those which are imaged with sufficient depth and resolution to allow us to detect faint satellite galaxies.  Furthermore, as the number of satellites is known to depend on the stellar mass of the host galaxy, we study only systems which had previously estimated stellar masses or multiple bands of photometry and a measured redshift, which enables us to estimate the stellar mass of the systems.  Figure \ref{fig:hostProps} contains a summary of the redshift, stellar mass and apparent magnitude distributions of the host galaxies. 

\subsection{Lens Host Galaxies}
In this work we focus mainly on measuring the properties of satellites around SLACS lens galaxies.  SLACS candidates were detected in Sloan spectroscopy by searching for spectra which were composed of early-type, low redshift lenses superimposed with a higher redshift emission-line galaxy. Lens status was then confirmed using either ACS or WFPC2 imaging to confirm the morphological characteristics expected in gravitational lensing \citep{Bol++04,Bol++05,Bol++06}.

We restrict our comparison to those confirmed SLACS early-type lenses with redshifts $z>0.1$ which have deep (multiple exposure) I$_{814}$ imaging in order to enable us to measure the satellite population using the same tools which we apply to COSMOS imaging.  We also require that the lenses were imaged with ACS rather than WFPC2, in order to ensure that there was a sufficient area of deep, high-resolution imaging around the lenses to allow for an accurate measurement of the satellite population. We excluded one system with detector artefacts very near the lens due to a saturated star, and one system which had four companions of comparable brightness in the field, making the detection of faint satellites impractical. The final sample contained 32 hosts. Host stellar masses, redshifts and photometry were obtained from \citet{Aug++10}. 

\subsection{Non-Lens Host Galaxies}
We select non-lens host galaxies from the COSMOS ACS survey. We study host galaxies with stellar masses greater than M$^*>10^{10.5}$M$_{\odot}$ as measured by \citet{Ilbert++10} using ground based photometry in conjunction with ACS imaging \citep{Aug++10}. Hosts are restricted be between redshifts 0.1$<$z$<1$ and to not be within R$_{200}$ of a more massive companion, where R$_{200}$ is estimated using the observed relationship between stellar mass and R$_{200}$ from \citep{Dutton++10}. This isolation requirement is included to ensure that we are not studying host galaxies which are themselves satellites of a more massive system.  We also use the strong gravitational lensing catalog by \citet{Faure++08} in order to exclude strong gravitational lenses from our sample. 

COSMOS host stellar masses are based on SExtractor's MAGAUTO \citep{Bertin++1996} from ACS imaging, unlike SLACS photometry which was performed using deVaucouleurs profile fitting. However, for bright galaxies, we found that this did not make a difference in COSMOS photometry \citep[see also][]{Benitez++04}. 

\begin{figure}
\centering
\includegraphics[scale=.4]{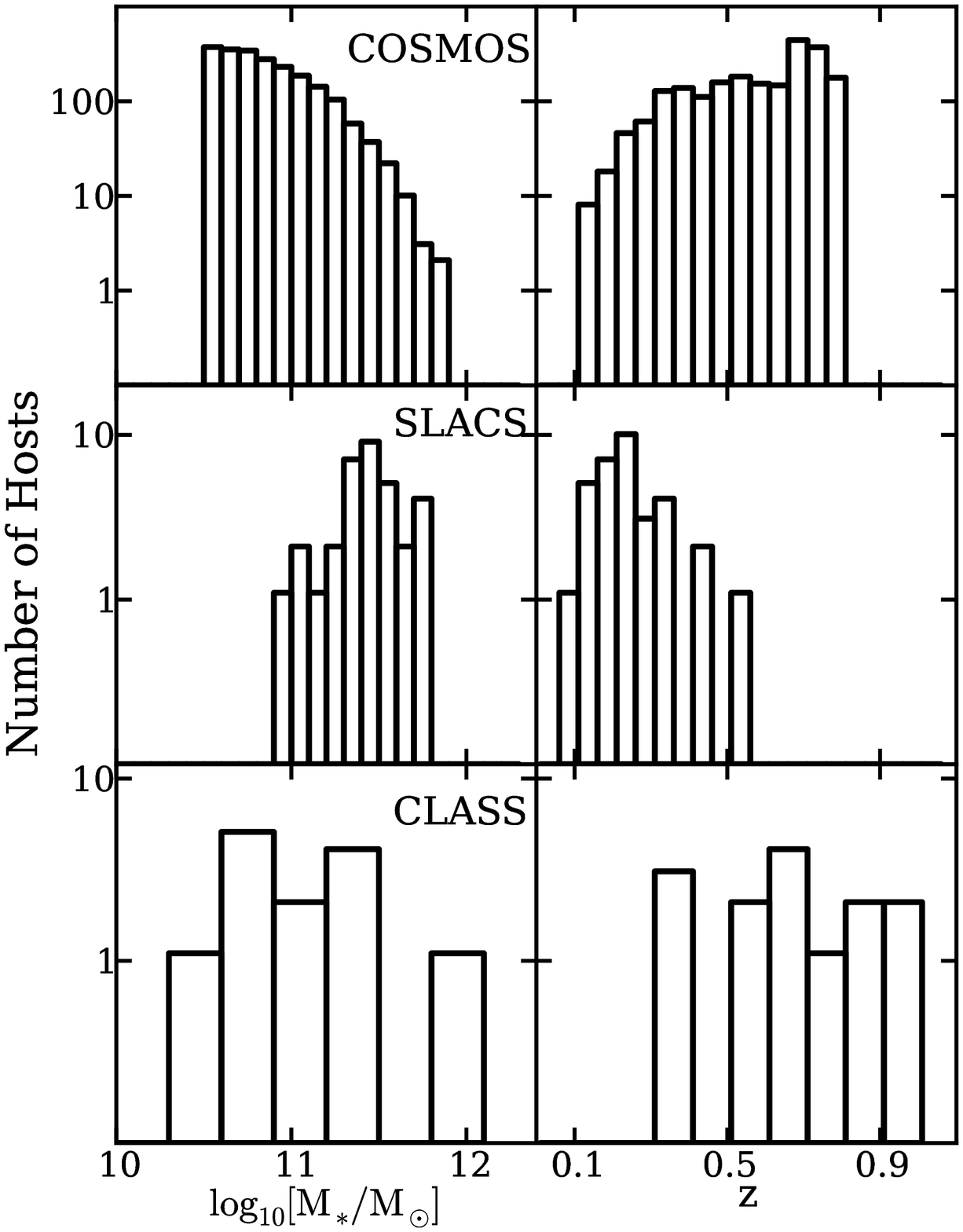}
\caption{Stellar mass and redshift distributions of host galaxies considered in this work. Lens galaxies are systematically more massive than non-lens galaxies. This is expected given that the lensing cross section increases as velocity dispersion to the fourth power.}
\label{fig:hostProps}
\end{figure}

\section{Companion Detection}
\label{sec:satelliteDetection}
We use the same method to detect companion objects around SLACS and COSMOS host galaxies as was used by N11 and N12. Briefly, the host galaxy light is modelled and removed from a cutout of $\sim 20$ r$_{\rm{eff}}$ around the I$_{814}$ images of host galaxies, using a modified version of the radial b-spline fitting code originally developed by \citet{Bol++05,Bol++06}. This method greatly facilitates the detection of nearby faint companions (see N11, N12).  Once the host light has been subtracted, we use SExtractor to detect nearby objects with parameters tuned to match the detection rate and photometry of the publicly available COSMOS ACS photometric catalogs \footnote{COSMOS photometric catalogs and surveys are available at http://irsa.ipac.caltech.edu/data/COSMOS/datasets.html}. We use masks of the lensed images created by the SLACS team \citep{Aug++10} to define a minimum radius outside of which we search for companion objects, thereby ensuring that lensed images were not included in the final catalog.

For COSMOS host galaxies, we compare our detections to those detections already in the COSMOS ACS photometric catalogs. New detections are added, while we replace the photometry for objects which were already in the photometric catalogs within $\sim 20$ r$_{\rm{eff}}$ of the host galaxies, with our own, thereby ensuring that the host light does not contaminate nearby objects.

For SLACS we detect objects in the full ACS images around the lens galaxies and then add in the new detections from the subtracted cutouts in the same way we do for COSMOS. For the SLACS images, bright stars and detector artefacts were masked by hand.

For both CLASS and SLACS hosts, we study all objects with I$_{814}<25$ magnitudes AB.

\section{Inference of the Number of Satellites Per Host}
\label{sec:statistics}
Our data consists only of single band photometry, so in order to study the properties of the satellite population we use the statistical method developed by N11 and N12, which we briefly review here. This method relies on the fact that the properties of the number density signal of background/foreground objects and satellite galaxies have different radial dependences relative to central host galaxies, allowing us to jointly infer the properties of the spatial distribution and the numbers of background/foreground objects and satellite galaxies around a sample of host galaxies. 

The statistical model can be summarised as follows; we infer the number of satellites per host within a fixed magnitude offset by modelling the spatial distribution of the number density signal near host galaxies as a combination of a homogenous background, with some mean value $\Sigma_{\rm{b,o}}$, and number counts slope N$_b(<\rm{m_{max}})\propto 10^{\alpha_{\rm{b}}(m_{\rm{h},i}+\rm{dm}-\rm{m_{max}})}$  and a satellite population which has a projected radial distribution given by $\rm{P}(r)\propto r^\gamma_{\rm p}$, and an average number of satellites per host N$_{\rm{s,o}}$. For simplicity we do not consider the angular distribution of satellites in this work.

In this work, we include two new additions to the model which allow for an accurate comparison between samples of host galaxies with different stellar mass distributions, and uncertain virial radii. We conclude by describing the priors used for the model. The likelihood function and a more detailed description of the statistics, including how we adjust for masked regions in the images can be found in Appendix A.

\subsection{Number of satellites as a function of host stellar mass}
In order to account for the differences in the stellar mass distributions of COSMOS and SLACS host galaxies (Figure \ref{fig:hostProps}), we infer the number of satellites as a continuous function of host stellar mass for the two samples.
\be
\rm{N}_{\rm {s}} \propto \rm{N}_{\rm {s,o}}(1+\log_{10}[\rm{M}^*_{\rm{host}}/\rm{M}\odot] - 11.4)^{\kappa_s}
\ee

Given that the amount of correlated structure is also a function of the host stellar mass, we also infer the number density of background objects as a function of stellar mass:
\be
\Sigma_b  = \left(\Sigma_{b,o} + \kappa_b(\log_{10}[\rm{M}^*_{\rm{host}}/\rm{M}\odot] - 11.4)\right) (1+z)^{\delta_{\rm{z,b}}}
\ee

This redshift evolution term in the foreground/background number density is not necessary for SLACS which covers a much smaller redshift range than the COSMOS hosts. We do not study redshift evolution in the number of satellites per host as there was no strong dependence apparent in N12.

\subsection{Uncertainty in the virial radius}
In order to account for variations in host halo mass and projected angular sizes due to redshift variations, it is important to choose an appropriate distance scale within which to study the satellite population. In N12 we scaled all distances by R$_{200}$ as estimated by the observed relationship between stellar and halo mass from \citet{Dutton++10}. This relationship becomes increasingly uncertain for high stellar masses \citep[e.g.][]{Behroozi++10}. 

The median stellar mass of the SLACS lens galaxies is $\log[\rm{M}^*_{\rm{host}}/\rm{M}\odot]=11.4$, which is approximately where the halo mass to stellar mass relation becomes very uncertain  \citep[e.g.][]{Behroozi++10, Leauthaud++11}. To account for this, for all hosts with stellar masses greater than $\log[\rm{M}^*_{\rm{host}}/\rm{M}\odot]>11.4$, the true virial radius is included as a model parameter, with mean and standard deviation given by \citet{Dutton++10}. We further restrict the inference to only allow for virial masses $\rm{M}_{\rm{vir}}<10^{14} M\odot$, as there are no clusters in either the COSMOS or SLACS samples, given group richnesses and lensing profiles. For our final results, we marginalise over the inferred virial radii. We note that the exact threshold where virial mass uncertainty is included does not significantly affect our results, as at lower stellar masses the stellar mass to virial mass uncertainty becomes smaller and the uncertainty due to the significantly smaller number of satellites per host dominates.

\subsection{Priors}
We use Gaussian priors on the parameters describing the number density of background/foreground objects ($\Sigma_{\rm{b,o}}$, $\kappa_b$ and $\delta_{\rm{z,b}}$). We obtain these priors by measuring the background properties in annuli between $1.0<r<1.5$ R$_{200}$ around the COSMOS hosts. 
For SLACS, we use priors with the same means as in COSMOS, but with broader standard deviations to allow for the fact that the COSMOS field is on average denser than pure-parallel fields \citep{Fassnacht++11}.  We do not allow for redshift variation in the background/foreground number density in SLACS due to the small sample size and narrow redshift range of the central galaxies.

For COSMOS satellites, we apply a prior on the slope of the projected radial profile of $\gamma_{\rm p}$ from N12. Values for mean and standard deviations for these priors are listed in Table \ref{tab:priors}. We infer the properties of SLACS satellites with and without this prior on $\gamma_{\rm p}$. 

We infer all parameters in bins of magnitude offset between companion and host galaxy ($\Delta \rm{m} = \rm{m}_{\rm{sat}}-\rm{m}_{\rm{host}}$).
\begin{table*}
\begin{center}
\scriptsize\begin{tabular}{llll}
\hline \hline
Parameter &  Description & COSMOS Prior &SLACS Prior\\
\hline 
N$_{\rm{s,o}}$ & Number of satellites for hosts with $\rm{M}^*= 10^{11.4}\rm{M}_{\odot}$&U(0,100)$^{a}$&U(0,100)\\ 
$\kappa_s$ &Dependence of satellite number on host stellar mass &U(0,60)&U(0,60)\\
$\gamma_{\rm{p}}$& Projected radial slope of the satellite number density &N(-1.1,0.3)$^{b}$ &U(-9,0)   \\ 
$\Sigma_{\rm{b,o}}$& Number of background/foreground objects per arcmin$^2$ with I$_{814}<25$ &N(47,0.3)&N(47,2) \\ 
$\alpha_{b}$& Slope of the background/foreground number counts as a function of magnitude &N(0.300,0.005) &N(0.3,0.1) \\ 
$\kappa_b$ &Dependence on the number of background/foreground objects on host stellar mass  &  N(1.5,0.3)&  N(1.5,0.3)\\
$\delta_{\rm {z, b}}$ &Dependence on the number of background/foreground objects with redshift.             &    N(0.03,0.02)  & ... \\
\hline \hline
\end{tabular}
\caption{Priors used in the inference of the number and spatial distribution of satellite galaxies around SLACS and COSMOS host galaxies.
(a) Uniform prior between (min,max),
(b) Normal prior with (mean, standard deviation) \label{tab:priors}}
\end{center}
\end{table*}

\section{Results}
\label{sec:results}
In Tables \ref{tab:SLACSresults} and \ref{tab:COSresults}, we present the results of our inference on the cumulative number of satellites per host galaxy as a function of stellar mass and host redshift, in two bins of magnitude offset between host and companion objects $\Delta \rm{m} = \rm{m}_{\rm{sat}}-\rm{m}_{\rm{host}}$.

\begin{table*}
\begin{center}
\scriptsize \begin{tabular}{cccccc|cc}
\hline
\hline
$\Delta$ m & N$_{\rm{s,o}}$ & $\kappa_s$ &$\gamma_{\rm{p}}$& $\Sigma_{b,o}$ & $\alpha_b$& N$_{\rm{s,o}}$ $^{a}$& $\kappa_s$ $^{a}$ \\
\hline	     
4 & 6$^{+3}_{-3}$ &2.1$^{+1}_{-0.7}$ & -0.7$^{+0.2}_{-0.3}$ & 47$^{+2}_{-2}$ & 0.38$^{+0.05}_{-0.04}$ & 3$^{+2}_{-2}$ &3$^{+1}_{-1}$ \\ 
5 & 9$^{+5}_{-3}$ & 2.4$^{+1}_{-0.8}$& -0.8$^{+0.2}_{-0.3}$ & 47$^{+2}_{-2}$ & 0.36$^{+0.04}_{-0.02}$&  7$^{+3}_{-3}$& 2.8$^{+1}_{-0.9}$  \\ 
6 & 20$^{+5}_{-4}$ &2.7$^{+0.5}_{-0.5}$& -0.7$^{+0.1}_{-0.1}$ & 47$^{+2}_{-2}$ & 0.39$^{+0.02}_{-0.02}$& 18$^{+4}_{-4}$& 2.7$^{+0.7}_{-0.6}$  \\ 
7 & 28$^{+7}_{-6}$ &2.7$^{+0.5}_{-0.5}$  & -1.0$^{+0.1}_{-0.2}$ & 46$^{+2}_{-2}$ & 0.39$^{+0.02}_{-0.02}$&27$^{+5}_{-5}$& 2.7$^{+0.5}_{-0.5}$   \\ 
8 & 37$^{+9}_{-7}$ &3.4$^{+0.7}_{-0.5}$& -1.0$^{+0.1}_{-0.2}$ & 42$^{+2}_{-2}$ & 0.39$^{+0.02}_{-0.02}$ & 38$^{+7}_{-7}$ &3.5$^{+0.6}_{-0.6}$ \\ 
\hline \hline
\end{tabular}

\caption{Summary of inference results for SLACS satellites, see Equations 1 and 2. Inferred between $0.03<R<0.5~ R_{200}$. (a) Inferred with a Gaussian prior on $\gamma_{\rm p}$ with mean -1.1 and standard deviation 0.3 \label{tab:SLACSresults}}
\end{center}
\end{table*}

\begin{table*}
\scriptsize\begin{tabular}{lllll}
\hline
\hline
$\Delta$ m &  N$_{\rm{s,o}}$ & $\kappa_{\rm{s}}$& $\gamma_{\rm{p}}$\\
\hline
2 & 0.8$^{+0.2}_{-0.2}$ & 1.4$^{+0.3}_{-0.3}$& -1.1$^{+0.1}_{-0.1}$  \\ 
3 & 1.7$^{+0.3}_{-0.3}$ &  1.7$^{+0.3}_{-0.3}$&-1.2$^{+0.1}_{-0.1}$  \\ 
4 & 2.9$^{+0.4}_{-0.4}$ & 1.6$^{+0.3}_{-0.2}$&-1.1$^{+0.1}_{-0.1}$  \\ 
5 & 4.2$^{+0.7}_{-0.8}$  & 1.8 $^{+0.4}_{-0.3}$ & -1.1$^{+0.1}_{-0.1}$\\ 
6 & 12$^{+1}_{-1}$     & 2.3$^{+0.4}_{-0.3}$ & -0.6$^{+0.1}_{-0.1}$ \\ 
7 & 20$^{+3}_{-3}$     &  1.8$^{+0.5}_{-0.5}$ & -0.8$^{+0.2}_{-0.2}$  \\ 
\hline
\hline
\end{tabular}
\caption{Summary of inference results for COSMOS satellites inferred between 0.03$<r<0.5~R_{200}$, see Equations 1 and 2. \label{tab:COSresults}}

\end{table*}

The results can be summarised as follows. For both COSMOS and SLACS hosts, we detect a significant dependence of the number of satellites per host on the host stellar mass, which is consistent between the samples.  These results are also consistent with the results of N12, in which we observed that the number of satellites per host at fixed $\Delta$ m  was significantly higher for hosts with $11<\log[\rm{M^*_{host}}/\rm{M}_\odot]<11.5$ than for hosts with $10.5<\log[\rm{M^*_{\rm{host}}}/\rm{M}_\odot]<11$. 


We find that the slope of the power-law radial profile of the satellite number density of satellites around SLACS lenses is consistent with that of COSMOS field galaxies and goes approximately as $\gamma_{\rm p} \sim -1$. As discussed by N11 and N12, this slope is approximately isothermal.  Given the range of host concentrations across our sample, and the fact that we are fitting only a single power-law to the radial profile of satellites, this result can be reproduced by a sum of Navarro Frank and White radial profiles \citep{NFW++1996} for the satellite number density, which is approximately the density profile which would be expected if satellite galaxies followed the density profile of the host halo, as predicted by simulations [e.g.][]\citep{Krav++10}.

Therefore, we perform the inference a second time, for SLACS galaxies using the same Gaussian prior on the projected radial profile as was used for COSMOS. These results are listed in the last two columns of Table \ref{tab:SLACSresults}. The Gaussian prior lowers the uncertainty in the inferred number of satellites, but it does not significantly affect the number of satellites per host. 

\begin{figure}
\centering
\includegraphics[scale=.4]{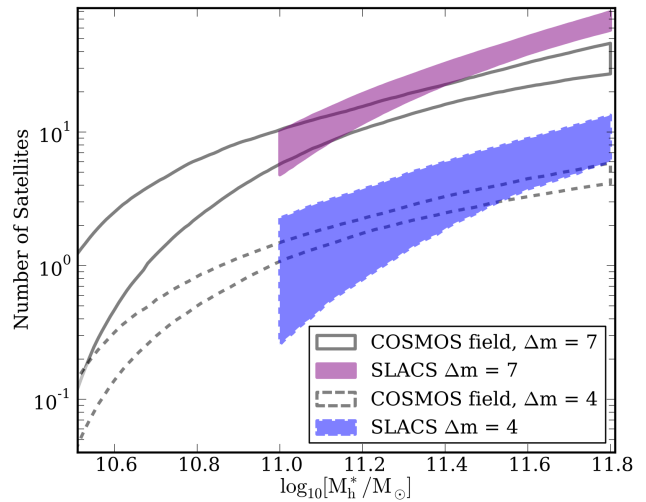}

\caption{The average cumulative number of satellites per host between 0.03$<R_{200}<$0.5, near COSMOS and SLACS hosts, for two values of $\Delta$m where all objects analysed are restricted to have $m_{\rm{obj}}-m_{\rm{host}}<\Delta$m . The filled curves show the one sigma confidence intervals for the inferred number of satellites per host as a continuous function of the host stellar mass. The inferred values for these functions are listed in Table \ref{tab:COSresults}. These results are inferred with a Gaussian prior on the projected slope of the satellite radial profile $\gamma_{\rm p}$ with mean $-1.1$ and standard deviation 0.3.}
\label{fig:slacscoscompare}
\end{figure}

In Figure \ref{fig:slacscoscompare}, we compare the inferred number of satellites for COSMOS and SLACS hosts using the same prior on $\gamma_{\rm p}$. The number of satellites as a function of host stellar mass is consistent at all stellar mass values, although there may be a small trend in a steeper slope for the SLACS satellites, the difference is significant only at the $\sim 1\sigma$ level. Note that although values of individual parameters are not always consistent for COSMOS and SLACS, the average number of satellites as a function of host stellar mass is, due to degeneracies between the parameters $\kappa_s$ and $\rm{N}_{\rm {s,o}}$, and the final number of satellites as a function of stellar mass. In Figure 3 we show the bivariate posterior probability distributions for $\kappa_s$ and $\rm{N}_{\rm {s,o}}$ for $\Delta$m $=4$ and $7$.

\begin{figure}
\centering
\includegraphics[scale=.4]{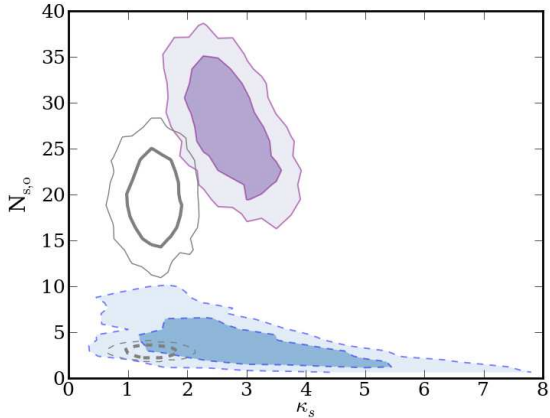} 
\caption{Comparison between posterior probabilities for the parameters $\kappa_s$ and N$_{\rm{s,o}}$ for $\Delta$m$=4$ and $7$, for SLACS and COSMOS host galaxies. The colours have the same meaning as in Figure 2. The darker inner contours and the lighter outer contours represent the 68 and 95 \% confidence intervals respectively.}

\end{figure}

\section{Comparison with CLASS}
\label{sec:CLASScompare}

Although the focus of this work is with the satellite population of SLACS lenses, it is interesting to see whether the CLASS lenses have a relatively higher rate of companion detection than COSMOS galaxies, given the stellar mass distribution of CLASS hosts, and the results for our inference on the dependence between satellite number and host stellar mass.

Gravitational lenses in the Cosmic Lens All Sky Survey \citep[CLASS,][]{Mye++03,Bro++03} were selected by searching for flat-spectrum radio sources with multiple spatial components. This selection is very distinct from the SLACS selection, as it is purely based on the detection of a gravitationally lensed radio source, and has no requirements for the detection of the deflector galaxy. For this reason, the CLASS galaxies span a much larger range in redshift space than SLACS galaxies, and they do not all have well measured photometry or redshifts. For this comparison we consider only CLASS hosts which have I$_{814}$ imaging with either WFPC2 or ACS \footnote{with the exception of 1422 which has I$_{791}$ imaging} and spectroscopic redshift measurements. 

We obtained stellar masses of the CLASS lenses by compiling photometry from the literature \citep{Lagattuta++10,Rusin++05,Mu++98}, and applying the stellar population synthesis modelling code by \citet{Aug++10}. Due to the non uniform methods of obtaining photometry across the different works, we assume a factor of $\sim$ 2 uncertainty in estimated stellar masses. We further restrict our sample to those lenses which have estimated stellar masses greater than M$^*/$M$_\odot>10^{10.4}$ in order to allow us to continue to use the functional form from Equation 1. 
The final comparison sample consists of 14 hosts \footnote{CLASS~B0128+437, MG0414+0534, CLASS~B0445+123, CLASS~B0631+519, CLASS~B0712+472, CLASS~B1030+074, CLASS~B1152+199, CLASS~B1422+231, CLASS~B1608+656, CLASS~B1933+503, CLASS~B1938+666, CLASS~B2045+265, CLASS~B2108+213, CLASS~B2319+051} with properties shown in Figure \ref{fig:hostProps}. 

We use companion object detection from \citet{Jackson++10} who used SExtractor to detect objects within 20 kpc of the CLASS deflectors, with an I$_{814}$ magnitude limit of 24.9. \citet{Jackson++10} visually excluded lensed images to ensure they were not counted as companion objects. After these considerations, \citet{Jackson++10} detects four companion objects within 20 kpc of the 14 CLASS host galaxies which have spectroscopic redshift measurements, and meet our stellar mass requirements.

Using the stellar mass distribution of the CLASS hosts, we apply the results from the previous section to estimate the number of companion objects we would have expected around COSMOS hosts with the same stellar mass distribution.  To do this, we drew randomly ten thousand times from the posterior probability distribution functions for $\kappa_s$,$\gamma_{\rm p}$ and N$_{\rm{s,o}}$, as well as from Gaussian distributions centred on the estimated stellar masses with standard deviations of 0.3 dex in order to account for uncertainty in the host stellar mass estimates. We used the values of $\gamma_{\rm{p}}$ to rescale the expected number of satellites to the smaller area studied by \citet{Jackson++10}.


 Assuming that the closest satellites could be found only as close as 10 kpc, to the host centres, due to obscuration by the host galaxy, a sample of 14 COSMOS hosts with the same stellar masses, and brightnesses as the CLASS hosts would have about $4\pm 2$ satellites, and approximately $4\pm 2$ background/foreground objects, depending on the assumed background/foreground number density around CLASS hosts. Given the small number statistics, this prediction is marginally consistent with the detection of four objects around CLASS hosts, however these numbers do not take into account obscuration by lensed images. 
 
 We can roughly account for obscuration by lensed images, by assuming that the lensed images are found in annuli at the lens Einstein radii, with radial width of 0$\farcs4$. For double image lenses, we assume that about one third of the annulus is blocked by lensed images, and for four image lenses, we assume the full annulus is blocked, although the exact fraction of blocking does not significantly affect the results. Using this approximate accounting for lensed image obscuration, we expect 2-3 satellite galaxies and 1-2 foreground/background objects to be detected around CLASS host galaxies. 
 
Although the main search radius in \citet{Jackson++10} was within 20 kpc, many of the companion objects around CLASS lenses were found even closer to the host galaxy, and brighter than the detection limit of 24.9. If we instead consider the number of objects between 4 and 14 kpc, within 3.3 magnitudes of the central galaxy magnitude, our model would predict about 2 satellites and 1 background/foreground object, including the effects of obscuration which are less relevant this close to the central galaxy, in comparison with the four companion objects \citet{Jackson++10} detected which met these criteria.
 
Therefore, taking into account the fact that the CLASS host galaxies are very massive, we find that they have approximately the same number of nearby companion objects as we observe around COSMOS host galaxies with comparable stellar masses. This result highlights the importance of considering host stellar masses when comparing samples of satellite galaxies.

\section{Summary}
\label{sec:Discussion}

We have measured the spatial distribution and number of satellites per host as a function of host stellar mass for SLACS lens galaxies and COSMOS field galaxies using host light subtraction, object detection and an updated version of the statistical analysis developed by \citet{Nierenberg++11,Nierenberg++12}. Furthermore, we compared the number of projected companion objects found within 20 kpc apertures of CLASS lens galaxies by \citet{Jackson++10} to what we would have expected around COSMOS field hosts with similar stellar masses

Our main results are summarised below:

\begin{enumerate}
\item We detect a significant population of luminous satellites around SLACS lens galaxies. Parametrising the spatial distribution of satellites as $P_{\rm{sat}}(r)\propto r^{\gamma_{\rm{p}}}$ we find $\gamma_{\rm{p}} \sim -0.8 \pm 0.2$, which is consistent with the spatial distribution of luminous satellites around COSMOS non-lens galaxies. 

\item Parametrising the number of satellites per host within a fixed magnitude offset from the host galaxy ($\Delta \rm{m} = \rm{m}_{\rm {sat}}-\rm{m}_{\rm{host}}$) to be $\rm{N}_{\rm {s}} \propto \rm{N}_{\rm {s,o}}(1+\rm{\log_{10}[\rm{M}^*_{\rm{host}}/\rm{M}_\odot]} - 11.4)^{\kappa_s}$, we find $\kappa_s$ to be $\sim 2.9 \pm$ 0.6 for SLACS satellites as measured for hosts between $10^{11}<\log_{10}[\rm{M}^*_{\rm{host}}/\rm{M}_\odot]<10^{11.8}$, and $\sim 1.8 \pm$ 0.5 for COSMOS satellites as measured between $10^{10.5}<\log_{10}[\rm{M}^*_{\rm{host}}/\rm{M}_\odot]<10^{11.8}$. Given degeneracies between $\kappa_s$ and N$_{\rm {s,o}}$, the number of satellites per host as a function of host stellar mass is consistent within the measurement uncertainties for all values of $\Delta$m.

\item Using the above results, we find that the number of close companions to CLASS lenses found by \citet{Jackson++10} is consistent with what we observe around COSMOS non-lens galaxies, taking the distribution of CLASS host stellar masses into account, as well as obscuration due to bright lensed images.

\end{enumerate}

From these results we conclude that the subhalo mass function measured from these strong gravitational lenses is representative of the global subhalo mass function for haloes of the same masses.
\section*{Acknowledgments}

A.M.N., D.O. and T.T.  acknowledge support from the Packard Foundations through
a Packard Research Fellowship. A.M.N. and D.O. also thank the Worster family for their research fellowship.
We sincerely thank A. Sonnenfeld for estimating stellar masses of CLASS lenses, and B. Kelly, C. Fassnacht, 
and N. Jackson, and the anonymous refferee for extremely helpful comments and discussions.
This work was based on observations made with the
NASA/ESA Hubble Space Telescope, and obtained from the Data Archive at
the Space Telescope Science Institute, which is operated by the
Association of Universities for Research in Astronomy, Inc., under
NASA contract NAS 5-26555.  These observations are associated with the
COSMOS and GOODS projects. 
\bibliographystyle{apj}
\bibliography{references}

\appendix
\section{Likelihood Function}
Here we describe how we infer the model parameters $\modpars$ (listed in Table 1), which describe the number and spatial distribution of satellite and background/foreground galaxies between 0.03 and 0.5 R$_{200}$ as a function of host stellar mass, within a fixed magnitude offset from the host galaxy ($ \rm{m}_{\rm{obj}}-\rm{m}_{\rm{host}}<\Delta \rm{m}$), given our data $\allDat$.

Using Bayes' theorem, the probability of the model parameters being true given the data is proportional to:
\be
\prob(\modpars|\allDat) \propto \prob(\allDat|\modpars)\prob(\modpars)
\ee
Where $\prob(\modpars)$ is the prior on the model parameters.

For the $j^{th}$ host in our sample, the data consist of a set of object radii relative to the host galaxy $\bs{r}_j$, the total number of objects around the host galaxy, $\numobs_j$, and the host stellar mass $\hmass$.
The likelihood function can be decomposed into a product over each host galaxy of the individual likelihoods :
\be
\prob(\modpars|\allDat) \propto \prod_j^{\numhosts}\prob(\data_j|\modpars,\hmass)\prob(\modpars)
\ee

For each host galaxy, the likelihood is the probability of measuring $\numobs_j$ times the product of the likelihoods for each object position given the model parameters.

\be
\prob(\data_j|\modpars,\hdat_j) = \prob(\numobs_j|\modpars)\prod_i\prob(r_i|\modpars,\hdat_j)
\ee

The first term on the right hand side is a Poisson probability comparing the model predicted number of objects with the observed number of objects

The model prediction for the number of objects is given by the sum of the model prediction for the number of satellite galaxies and the number of background/foreground objects, where the number of background/foreground objects is given by:
\be
N_b = A\Sigma_b
\ee
Where $\Sigma_b$ is defined in Equation 1, and $A$ is the area (in arcminutes) in which objects were observable for a given host.

The model prediction for the observable number of satellites $N'_{s,j}$, for the $j^{th}$ host is given by:
\be
N'_{s,j} = N_{s,j} \frac{\int^{r_{\rm{max},j}}_{r_{\rm{min},j}} r^{\gamma_{\rm p}} f_j(r)  r dr}{\int^{0.5}_{0.03} r^{\gamma_{\rm p}} r dr}
\ee

Where $N_{s,j}$ is the model prediction for the number of satellites per host between 0.03 and 0.5 $R_{200,j}$, as defined in Equation 2,  $f_j(r)$ is the fraction of observable area as a function of radius, and $r_{\rm{min},j}$ and $r_{\rm{max},j}$ are the minimum and maximum observable radii respectively in units of  $R_{200,j}$. Note that this is where uncertainty in the virial radius enters in to the likelihood function. For hosts with stellar masses greater than $10^{11.4}$ M$_\odot$, the virial radius is an additional model parameter.

The second term on the right hand side of Equation A3 is the likelihood of the observed objects being at their observed positions given the model parameters. As we do not know whether a certain object is a satellite or a foreground/background object, the total probability of it appearing at a given position is the sum of the probability of it being there if it was a satellite, and the probability of it being there if it was a background/foreground object, weighted by the relative probabilities of an object being a satellite or a background/foreground object given the model parameters:
\begin{equation}
\begin{array}{ll}
\prob(r_i|\modpars,\hdat_j) =&  \prob(r_i|\modpars,\hdat_j,S)\prob(S|\modpars,\hdat_j) 
\\ &+\prob(r_i|\modpars,\hdat_j,B)\prob(B|\modpars,\hdat_j)
\end{array}
\end{equation}

The terms $\prob(S|\modpars,\hdat_j)$, and $\prob(B|\modpars,\hdat_j)$ are the relative probabilities of something being a satellite or a background/foreground object given the model parameters. For example:
\be
\prob(S|\modpars,\hdat_j)= \frac{N'_{s,j}}{N'_{s,j}+N_{b,j}}
\ee
The term $\prob(B|\modpars,\hdat_j)$ is defined analogously.

The probability of observing a satellite at position $r_i$ is given by:
\be
\prob(r_i|\modpars,\hdat_j,S)  = \frac{r_i^{\gamma_{\rm p}+1} f_j(r)}{\int^{r_{\rm{max},j}}_{r_{\rm{min},j}} r^{\gamma_{\rm p}}f_j(r) r dr}
\ee

The probability of observing a background/foreground object at position $r_i$ is:
\be
\prob(r_i|\modpars,\hdat_j,B) = \frac{r_i f_j(r)}{\int^{r_{\rm{max},j}}_{r_{\rm{min},j}} f_j(r) r dr}
\ee

\label{lastpage}
\bsp
\end{document}